# L'utilisation des splines bidimensionnels pour l'estimation de lois de maintien en arrêt de travail

**Frédéric P**LANCHET[*]   **Pascal W**INTER[α]

**ISFA – Université Lyon 1** [β]
**Winter & Associés** [γ]

RESUME

L'objectif de ce travail est de proposer un modèle d'ajustement paramétrique bi-dimensionnel pour des lois de maintien en incapacité de travail. La méthode proposée s'appuie sur des splines en dimension 2 ; elle est appliquée à un portefeuille réel, et le barème de provisionnement qui en résulte est comparé au barème de référence du BCAC.

MOTS-CLEFS :   Arrêt de travail, Kaplan-Meier, splines, estimation.

ABSTRACT

The aim of this paper is to propose an operational two-dimensional parametric adjustment for laws of maintenance in disability. The method suggested rests on splines in dimension 2; it is applied to a real data set, and the scale of reserving which results from it is compared with the scale of reference of the BCAC.

---

[*] Frédéric Planchet est professeur associé de Finance et d'Assurance à l'ISFA (Université Lyon 1 – France) et actuaire associé chez Winter & Associés. Contact : fplanchet@winter-associes.fr.
[α] Pascal Winter est étudiant à l'ISFA. Contact : pwinter@lynxial.fr.
[β] Institut de Science Financière et d'Assurances (ISFA) - 50, avenue Tony Garnier 69366 Lyon Cedex 07.
[γ] Winter & Associés - 9, rue Beaujon 75008 Paris et 18, avenue Félix Faure 69007 Lyon.

# 1. Introduction

L'estimation de lois de maintien d'expérience en arrêt de travail constitue un pré-requis incontournable pour l'évaluation de provisions mathématiques sur des portefeuilles dont le comportement diffère *a priori* de la norme proposée par le BCAC (*cf.* l'article A335-1 du Code des Assurances).

L'expérience montre que le taux de sortie de l'arrêt de travail dépend notamment, à population fixée, de l'âge au début de l'arrêt et de l'ancienneté de cet arrêt. Cette observation conduit à élaborer des tables doublement indicées, par l'âge à l'entrée d'une part et par l'ancienneté d'autre part. C'est ce type de présentation qui est par exemple retenu par le BCAC dans les tables réglementaires qu'il propose.

Cependant, les méthodes statistiques couramment utilisées pour construire des lois de maintien sont essentiellement mono-dimensionnelles : c'est ainsi le cas pour l'estimation des taux bruts via l'estimateur de Kaplan-Meier, qui est à certains égards le meilleur en présence de censure (*cf.* DROESBEKE et al. [1989]), et également pour les méthodes d'ajustement paramétrique classiques, de type Makeham, logit ou splines (*cf.* PLANCHET [2004]).

En ce qui concerne l'estimation des taux bruts, des travaux s'intéressent à la généralisation de l'estimateur de Kaplan-Meier en dimension supérieure à un ; plus précisément, ces travaux concernent l'estimation de Kaplan-Meier en présence d'hétérogénéité (*cf.* XIE et LIU [2000]). L'estimation des taux bruts pour chaque âge à l'entrée par Kaplan-Meier classique reste toutefois actuellement la méthode la plus robuste pour produire les taux bruts. Au surplus, la perte d'information que représente la non prise en compte de la loi conjointe selon les deux dimensions du problème est faible et peu pénalisante en pratique.

Lorsque l'on souhaite ajuster les taux bruts ainsi obtenus (voir par exemple LONDON [1985] pour les motivations d'un ajustement), l'approche usuelle consistant à « recoller » des tables uni-dimensionnelles construites pour chaque âge à l'entrée possible (décrite par exemple dans GAUMET [2001]) rend délicate la prise en compte de la dynamique des taux de sortie en fonction de l'âge à l'entrée.

Parallèlement, des approches développées depuis quelques années dans le cadre de la détermination de tables de mortalité prospectives proposent des modèles naturellement bi-dimensionnels : il s'agit notamment du modèle de Lee-Carter (*cf.* LEE et CARTER [1992] et LEE [2000]) et de ses variantes de type log-Poisson proposées par BROUHNS et al. [2002], ou d'autres approches comme les modèles de type géostatistique de DEBON et al. [2004].

Dans le but de construire des tables de maintien en arrêt de travail avec des méthodes bi-dimensionnelles, *a priori* mieux adaptées au contexte que l'approche usuelle rappelée ci-dessus, on peut tenter d'adapter les modèles de type Lee-Carter au contexte de l'arrêt de travail, en faisant jouer à l'âge le rôle de l'ancienneté et à l'année calendaire le rôle de l'âge à l'entrée.

Toutefois les travaux de LELIEUR [2005] montrent que dès lors que l'on sort du contexte de très grands échantillons dans lesquels ces méthodes ont été développées pour travailler sur des portefeuilles nécessairement de taille plus réduite, une instabilité des coefficients liées aux fluctuations d'échantillonnage et à une certaine « sur-paramétrisation » apparaît et rend ces modèles inopérants, ou du moins peu efficaces. Cette difficulté est contournée dans le cas de la mortalité en réduisant le nombre de paramètres du modèle au travers d'une spécification de la forme fonctionnelle de la dépendance en fonction de l'âge et de l'année des coefficients concernés (*cf.* LELIEUR [2005]). La transposition au cas de l'arrêt de travail, effectuée par WINTER [2005], s'avère globalement peu satisfaisante, les modèles de Lee-Carter ou log-Poisson s'avérant peu à même de bien prendre en compte la forme particulière des taux de sortie pour ce risque.

L'objectif du présent travail est de présenter une nouvelle méthode utilisant les splines en dimension 2. L'utilisation des splines dimension 2 est initialement née dans le contexte d'applications industrielles, et est devenu une méthode classique dans ce domaine : on pourra notamment consulter RISLER [1991] et DE BOOR [1978]. On pourra se reporter à BESSE et CARDOT [2001] pour une synthèse des méthodes statistiques de lissage dans les espaces fonctionnels. L'utilisation de splines bi-dimensionnels dans le domaine de la construction de lois de maintien constitue une innovation, même si en dimension un il s'agit d'un outil souvent utilisé.

Après avoir exposé l'approche par splines, on présente une application sur le cas d'un portefeuille d'arrêt de travail pour le risque incapacité ; l'application à la construction d'un barème de provisionnement de ce risque est ensuite présentée.

Le présent travail est inspiré d'un travail de recherche effectué dans le cadre de l'Institut de Science Financière et d'Assurances de l'Université Lyon 1 par WINTER [2005].

## 2. L'ajustement par splines

### 2.1. Présentation générale

L'ajustement à une loi continue suppose implicitement que la courbe des taux de sortie peut être représentée sur toute la plage d'ancienneté considérée par une seule fonction paramétrique. En pratique, du fait par exemple de ruptures dans l'évolution des taux bruts, cette condition apparaît assez restrictive dès lors que l'on veut rester dans le cadre d'une fonction « simple ».

L'idée du lissage par splines est de découper la plage de la fonction à ajuster en sous-intervalles, puis d'ajuster sur chaque sous-intervalle une fonction simple, en prenant des précautions pour le raccordement aux points de jonction. Un découpage bien choisi doit en effet permettre d'utiliser sur chaque sous-intervalle une fonction sensiblement plus simple que la fonction qu'il aurait fallu ajuster globalement.

Les polynômes sont des fonctions simples et peuvent à ce titre être utilisés pour construire des lissages par spline ; en pratique, on considère en général des polynômes de degré 3 qui vont nous permettre de construire des splines cubiques. Le raccordement de ces arcs se fera en imposant aux points de jonction la continuité ainsi que l'égalité des pentes et des courbures.

La formalisation mathématique de cette approche en dimension un est relativement simple ; toutefois, en dimension deux elle s'avère un peu plus délicate et nécessite un minimum de formalisme, détaillé ci-après.

On peut simplement retenir que l'idée est de fixer arbitrairement des nœuds pour subdiviser la surface à ajuster surface en zones. Ensuite, on ajuste un polynôme de degré 3 (en dimension 2) à chaque subdivision, en utilisant un critère de type moindres carrés. Enfin, pour des raisons évidentes de continuité et de régularité, ce « patchwork » de polynômes de degré 3 est contraint à être de classe $C^2$ (ou $C^1$).

## 2.2. Mise en œuvre dans le cas de lois de maintien bi-dimensionnelles

Soit $P(t,x)$ un polynôme de degré $n$ en fonction de $t$ et $x$. Il est entièrement défini par la donnée de $(n+1)^2$ paramètres :

$$P(t,x) = \sum_{(i,j) \in \{0,\ldots,n\}^2} a_{ij} t^i x^j \,, \tag{1}$$

On note à présent $A = (a_{ij})_{(i,j) \in \{0,\ldots,n\}^2}$ la matrice des coefficients du polynôme $P(t,x)$. On note également $T_n = \begin{pmatrix} 1 \\ \vdots \\ t^n \end{pmatrix}$ et $X_n = \begin{pmatrix} 1 \\ \vdots \\ x^n \end{pmatrix}$ et $X'_n = \begin{pmatrix} 0 \\ 1 \\ \vdots \\ nx^{n-1} \end{pmatrix}$ le vecteur dérivé terme à terme. On obtient ainsi la représentation matricielle du polynôme :

$${}^t T_n . A . X_n = P(t,x) \,. \tag{2}$$

### 2.2.1. Calculs préliminaires

Les conditions de régularité nécessitent le calcul des dérivées partielles premières et secondes de $P(t,x)$. On obtient aisément :

$$\begin{aligned}
\frac{\partial P}{\partial t}(t,x) &= \sum_{(i,j) \in \{1,\ldots,n\}*\{0,\ldots,n\}} i a_{ij} t^{i-1} x^j & \frac{\partial P}{\partial x}(t,x) &= \sum_{(i,j) \in \{0,\ldots,n\}*\{1,\ldots,n\}} j a_{ij} t^i x^{j-1} \\
\frac{\partial^2 P}{\partial t^2}(t,x) &= \sum_{(i,j) \in \{2,\ldots,n\}*\{0,\ldots,n\}} i(i-1) a_{ij} t^{i-2} x^j & \frac{\partial^2 P}{\partial x^2}(t,x) &= \sum_{(i,j) \in \{0,\ldots,n\}*\{2,\ldots,n\}} j(j-1) a_{ij} t^i x^{j-2} \\
\frac{\partial^2 P}{\partial t \partial x}(t,x) &= \sum_{(i,j) \in \{1,\ldots,n\}*\{1,\ldots,n\}} ij a_{ij} t^{i-1} x^{j-1}
\end{aligned} \tag{3}$$

Les deux premières sommes comportent $n(n+1)$ termes, les deux suivantes $(n-1)(n+1)$ et la dernière $n^2$ termes. Ces expressions s'écrivent simplement sous forme matricielle :

$$\frac{\partial P}{\partial t}(t,x) = {}^\tau T'_n . A . X_n \qquad \frac{\partial P}{\partial x}(t,x) = {}^\tau T_n . A . X'_n$$

$$\frac{\partial^2 P}{\partial t^2}(t,x) = {}^\tau T''_n . A . X_n \qquad \frac{\partial^2 P}{\partial x^2}(t,x) = {}^\tau T_n . A . X''_n \qquad (4)$$

$$\frac{\partial^2 P}{\partial t \partial x}(t,x) = {}^\tau T'_n . A . X'_n$$

### 2.2.2. Nœuds

On appelle nœuds les points intérieurs de la table choisis pour définir les zones sur lesquelles un unique polynôme va être appliqué. Ces points doivent choisis aussi judicieusement que possible, en fonction de la forme et l'allure générale de la surface à représenter.

Par ailleurs, le choix du nombre de nœuds influe grandement sur les résultats obtenus. Un faible nombre de nœuds conduit à un lissage peu fidèle, un nombre élevé engendre quand à lui un lissage trop fidèle et donc irrégulier.

Pour plus de simplicité, on ne définit ici que des droites horizontales et verticales, distinctes des bords de la table. Les intersections de ces droites forment alors les nœuds.

Si on définit $h$ droites horizontales, et $v$ droites verticales, on obtient $hv$ nœuds, et la table est subdivisée en $(h+1)(v+1)$ parties. On indice alors ces parties avec $k = 0,\ldots,h$ et $l = 0,\ldots,v$, ce qui conduit à $(h+1)(v+1)$ polynômes à estimer et à recoller.

Dans notre problématique d'ajustement, la fonction brute est connue et donc représentable graphiquement. Une manière de choisir les nœuds est alors l'analyse préalable des données brutes. Les grandes tendances pourront être conservées avec un maillage épais ; néanmoins, certaines irrégularités (les pics en particulier) seront effacées. Pour éviter ceci, il faut encadrer ces irrégularités par la plus petite maille possible.

Une approche possible est donc de considérer que plus la variation de la fonction à ajuster est brutale, plus le maillage doit être petit si l'on veut conserver ces variations.

DE BOOR [1978] signale un point sensible : si une irrégularité se présente diagonalement par rapport aux composantes, les nœuds scindant la surface brute en rectangles peuvent ne pas être adaptés. On peut alors considérer un découpage de la surface à lisser en triangles.

Cette approche empirique de détermination des nœuds peut être remplacée par une approche de sélection automatisée. Dans la présente étude, le choix des nœuds est réalisé automatiquement par l'algorithme d'optimisation retenu (*cf. infra*). Le critère s'appuie sur un résultat prouvé par DE BOOR et FIX [1973] fournissant une majoration de la distance entre la fonction spline et la fonction à ajuster.

Cette inégalité permet alors de définir un critère de choix d'un jeu de nœuds efficace en recherchant la stratégie qui minimise la majoration. Le détail de l'approche est décrit dans DE BOOR et FIX [1973].

### 2.2.3. Conditions Limites

Pour des raisons de continuité, le raccordement entre les splines est soumis à des contraintes sur les dérivés des polynômes. Rappelons que dans la représentation matricielle du problème chaque ligne représente une ancienneté, chaque colonne un âge.

**Contraintes horizontales : ancienneté fixe**

Si $P^1$ et $P^2$ sont adjacents horizontalement, alors leurs valeurs et leurs dérivées partielles premières et secondes en fonction de $t$ doivent être égale, ceci pour tout âge variant sur l'ensemble de définition induit par le spline. En notant $(a_{i,j})$ et $(b_{i,j})$ les paramètres des splines adjacents, ces contraintes s'écrivent :

$$\sum_{j=0}^{n}\sum_{i=0}^{n} a_{ij} t^i x^j = \sum_{j=0}^{n}\sum_{i=0}^{n} b_{ij} t^i x^j , \tag{5}$$

et

$$\sum_{j=0}^{n}\sum_{i=1}^{n} i a_{ij} t^{i-1} x^j = \sum_{j=0}^{n}\sum_{i=1}^{n} i b_{ij} t^{i-1} x^j . \tag{6}$$

avec $t$ fixé et pour tout $x$ variant dans l'ensemble de définition du spline. Cela conduit à écrire :

$$\begin{aligned} & \sum_{j=0}^{n} x^j \sum_{i=0}^{n} (a_{ij} - b_{ij}) t^i = 0 \\ \Leftrightarrow & \sum_{i=0}^{n} (a_{ij} - b_{ij}) t^i = 0 \quad \forall j = 0,\ldots,n \quad \Leftrightarrow \quad {}^\tau T_n (A - B) = (0,\ldots,0) \end{aligned} \tag{7}$$

,

et

$$\begin{aligned} & \sum_{j=0}^{n} x^j \sum_{i=1}^{n} (a_{ij} - b_{ij}) i t^{i-1} = 0 \\ \Leftrightarrow & \sum_{i=1}^{n} (a_{ij} - b_{ij}) i t^i = 0 \quad \forall j = 0,\ldots,n \quad \Leftrightarrow \quad {}^\tau T_n' (A - B) = (0,\ldots,0) \end{aligned} \tag{8}$$

Si $t \neq 0$, ces équations forment un système libre de $2(n+1)$ équations avec $2(n+1)^2$ paramètres.

**Contraintes verticales : âge fixe**

De même, si $P^1$ et $P^2$ sont adjacents horizontalement, alors leurs valeurs et leurs dérivées partielles premières et secondes en fonction de $x$ doivent être égale, ceci pour toute ancienneté variant sur l'ensemble de définition induit par le spline. Avec un raisonnement similaire à celui décrit *supra*, on obtient le système :

$$\sum_{j=0}^{n}(a_{ij}-b_{ij})x^j = 0 \quad \forall i=0,\ldots,n \quad \Leftrightarrow \quad (A-B)X_n = {}^\tau(0,\ldots,0)$$

$$\sum_{j=1}^{n}(a_{ij}-b_{ij})jt^j = 0 \quad \forall i=0,\ldots,n \quad \Leftrightarrow \quad (A-B)X_h = {}^\tau(0,\ldots,0)$$

(9)

**Contraintes aux nœuds : âge et ancienneté fixes**

On cumule les deux contraintes précédentes. Par ailleurs, en un nœud, il se trouve non plus deux mais quatre polynômes adjacents.

**Contraintes en bord de table**

En bord de table, il n'y a pas de contraintes de régularité, puisque le polynôme n'est pas raccordé à un autre.

### 2.2.4. Choix du critère d'optimisation

On combine un critère de fidélité et un critère de régularité, comme décrit *supra*.

**Critère de fidélité**

Ici, *SP* représente la surface des splines raccordés. Les polynômes ajustés doivent être le plus fidèle possible à la surface brute. Pour ceci, on utilise un critère des moindres carrés pour chaque spline :

$$\sum_{x}\sum_{t}\omega_{x,t}\bigl(SP(t,x)-q_{x,t}\bigr)^2 \quad (10)$$

où $\omega_{x,t}$ est le poids associé au point $q_{x,t}$. Notons que la sommation doit inclure les bornes, c'est-à-dire les points se trouvant sur les droites définissant l'emplacement des nœuds.

**Critère de régularité**

Comme dans le lissage de Whitaker-Henderson, on introduit un critère permettant de contrôler la régularité du spline. La fidélité à elle seule n'est pas suffisante pour l'obtention d'une courbe régulière. En effet, outre le choix du nombre de nœuds, plus une courbe est fidèle aux données brutes, moins elle sera régulière. Le meilleur moyen de contrôler la régularité est de minimiser la dérivée seconde des polynômes. Avec la norme $L^2$, cette contrainte s'écrit :

$$\iint \lambda(s,y)\bigl|D^2 P(s,y)\bigr|^2 ds\,dy \quad (11)$$

où $\lambda(x,t)$ est une fonction continue par morceaux, valant le poids associé au point $q_{x,t}$ pour la régularité et $D^2P(s, y)$ la matrice Hessienne[1] de la fonction polynomiale par morceau calculé au point $(s, y)$.

## 2.2.5. Résolution

Le système à résoudre combine finalement la fidélité aux données brutes, la régularité de la surface ajustée et les contraintes aux nœuds. On combine les deux premières contraintes pour former la fonction à minimiser. La troisième constituera les contraintes du problème d'optimisation. On obtient alors le critère d'optimisation suivant :

$$\alpha \sum_x \sum_t \omega_{x,t} \left( SP(t,x) - q_{x,t} \right)^2 + (1-\alpha) \iint \lambda(s,y) \left| D^2 P(s,y) \right|^2 ds\, dy \qquad (12)$$

Le paramètre $\alpha \in [0,1]$ permet de privilégier soit la régularité, soit la fidélité. Les contraintes aux nœuds s'expriment sous la forme du système linéaire suivant :

$$\forall (k,l) \in [\![1,h]\!] \times [\![1,v]\!] \quad \begin{cases} {}^\tau T_n(t^*_{k,l}).(P_{k,l} - P_{k-1,l}) = (0,\ldots,0) \\ {}^\tau T'_n(t^*_{k,l}).(P_{k,l} - P_{k-1,l}) = (0,\ldots,0) \\ (P_{k,l} - P_{k,l-1}).X_n(x^*_{k,l}) = {}^\tau(0,\ldots,0) \\ (P_{k,l} - P_{k,l-1}).X'_n(x^*_{k,l}) = {}^\tau(0,\ldots,0) \end{cases} \qquad (13)$$

où $P_{k,l}$ représente la matrice des coefficients du polynôme $P_{k,l}$. Les points $t^*_{k,l}$ et $x^*_{k,l}$ représentent les abscisses et ordonnées du nœud correspondant à l'intersection de la (k-1)$^{\text{ème}}$ droite verticale et de la (l-1)$^{\text{ème}}$ droite horizontale (rappelons que les droites sont indicées de 1 à $k$ (resp. 1 à $l$), et qu'elles sont distinctes des bords ; on a donc $(h+2)(v+2)$ nœuds). Pour des polynômes de degré $n$, on a alors $4hv(n+1)$ équations pour $(h+1)(v+1)(n+1)$ paramètres.

Ce système décrit une condition de régularité $C^1$ pour la PP-Surface estimée. Il est facilement généralisable à une condition $C^2$ : il faut alors ajouter 2 équations aux contraintes présentées ci-dessus :

$${}^\tau T''_n(t^*_{k,l}).(P_{k,l} - P_{k-1,l}) = (0,\ldots,0) \text{ et } (P_{k,l} - P_{k,l-1}).X''_n(x^*_{k,l}) = {}^\tau(0,\ldots,0) \qquad (14)$$

C'est ce qui sera retenu dans la suite.

---

[1] Il s'agit de la matrice carrée composée des dérivées secondes de *P*.

## 2.3. L'approche de DE BOOR [1978]

En pratique l'introduction simplifiée ci-dessus est formalisée au travers des B-splines, qui fournissent un cadre de travail particulièrement bien adapté pour la description des problèmes d'ajustement à des fonctions splines.

Dans son ouvrage sur les splines, DE BOOR détaille ce cadre de travail. Les deux notions principales sont la différence divisée $k^{\text{ème}}$ et les B-splines, qui permettent d'exprimer les « pp functions » (piecewise polynomial functions). La généralisation de l'approche en dimension deux se fait dans un second temps à l'aide des produits tensoriels.

Le problème traité ici est suffisamment simple pour se passer de ce formalisme, toutefois indispensable pour des questions plus complexes.

# 3. Application

Afin d'illustrer la pertinence de la méthode proposée, nous l'avons mise en œuvre sur un portefeuille, dans l'optique de proposer une table d'expérience certifiable. Les aspects en lien avec la certification, et notamment la justification du caractère prudent de la table d'expérience obtenue, ne sont pas abordés ici, notre objectif étant d'illustrer les étapes de la mise en œuvre des splines. Le lecteur intéressé par ces aspects pourra se reporter à PLANCHET [2004].

## 3.1. Présentation des données

La constitution de la base de données constitue en soi un travail délicat et décisif, car il conditionne la qualité et la robustesse des estimations effectuées ensuite ; cette étape n'est pas abordée ici, et nous nous contentons de fournir un bref descriptif des données brutes après mise en forme.

Le portefeuille utilisé contient environ 170 000 arrêts, observés sur quatre ans. Il s'agit d'arrêts d'une durée supérieure à onze jours, du fait des dispositions de la convention collective concernée. Dans la suite on étudiera donc des lois conditionnelles au fait que la durée de l'arrêt est supérieure (strictement) à 10 jours. Les données sont également partiellement censurées à droite, mais le taux de censure reste faible (et systématiquement inférieur à 10%). L'information transmise permet d'effectuer les calcul de Kaplan-Meier avec un pas quotidien.

Au global l'étude porte sur environ 615 000 arrêts, dont 22 500 censurés à droite, soit 3,7% de l'effectif. La moyenne globale de l'échantillon n'est pas estimable aisément, à cause de la présence des censures. Cependant, la durée moyenne d'arrêt pour les individus non censurés est de 88 jours (plus 10 jours de troncature droite si on les compte depuis le début) et l'écart-type de 176 jours. De même, la durée moyenne d'observation pour les individus censurés est de 328 jours avec 10 jours en option, et l'écart type vaut 284 jours.

On porte une attention particulière aux grandeurs suivantes, calculées âge par âge : effectifs d'arrêts observés, proportions d'arrêts non censurés (rapportées sur les effectifs globaux) et nombre d'arrêts ayant une durée supérieure à 1095 jours ; on obtient le graphe suivant :

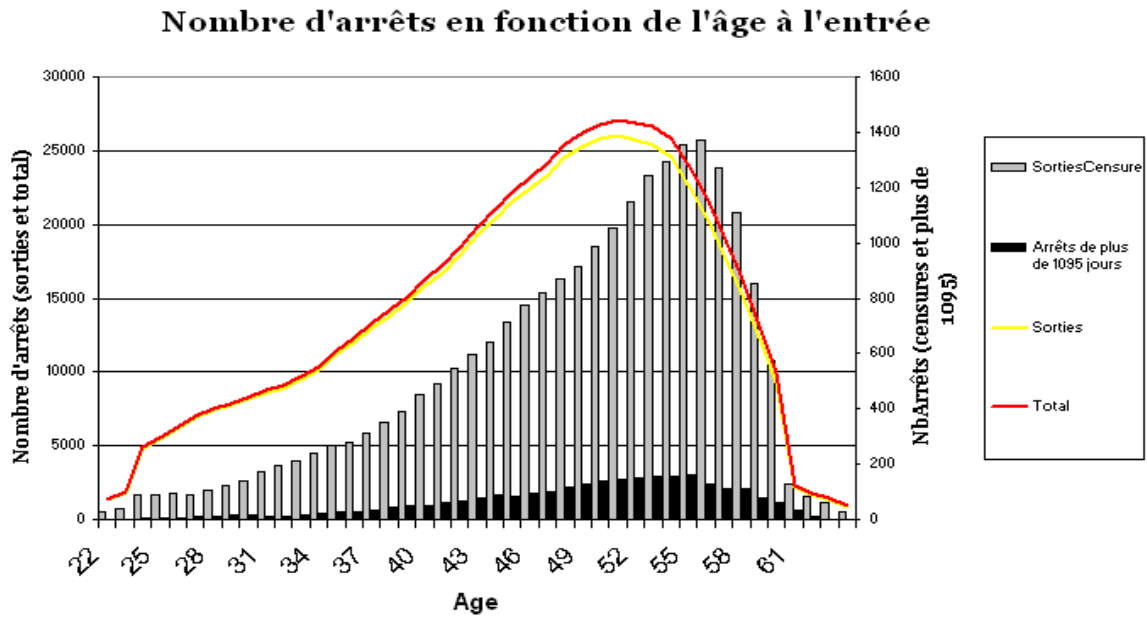

Fig. 1 :   *Analyse du nombre d'arrêt en fonction de l'âge à l'entrée*

## 3.2. Estimation des taux bruts

L'estimateur de Kaplan-Meier est calculé âge par âge :

$$\forall t \in \{AncMin,......, AncMax\} \quad \hat{S}_x(t) = \prod_{T_i \leq t}\left(1 - \frac{d_x(T_i)}{n_x(T_i)}\right) \quad (15)$$

avec $n_x(t)$ l'échantillon sous risque à l'âge $x$ juste avant le jour $t$ et $d_x(t)$ le nombre de sorties d'incapacité pour l'âge $x$ et l'instant $T_i$ (par reprise d'activité, passage en invalidité, décès ou fin de garanties).

En pratique, l'estimateur de Kaplan Meier donne lieu à un calcul que lorsqu'une personne sort de l'état ($T_i$). Par ailleurs, en toute rigueur, il n'y a pas d'ex aequo, car la probabilité que deux personnes sortent exactement au même instant est nulle presque sûrement. Néanmoins, comme les observations dont nous disposons sont volumineuses, et que, la durée des arrêts est comptée en jours (observations discrètes), on est confronté à la présence d'ex aequo. On utilise donc la formule approchée suivante :

$$\forall t \in \{AncMin......AncMax\} \quad \hat{S}_x(t) = \prod_{i=AncMin}^{t}\left(1 - \frac{d_x(i)}{n_x(i)}\right) \quad (16)$$

On obtient ainsi une estimation de la fonction de survie journalière pour tous les âges, présentée ci-après :

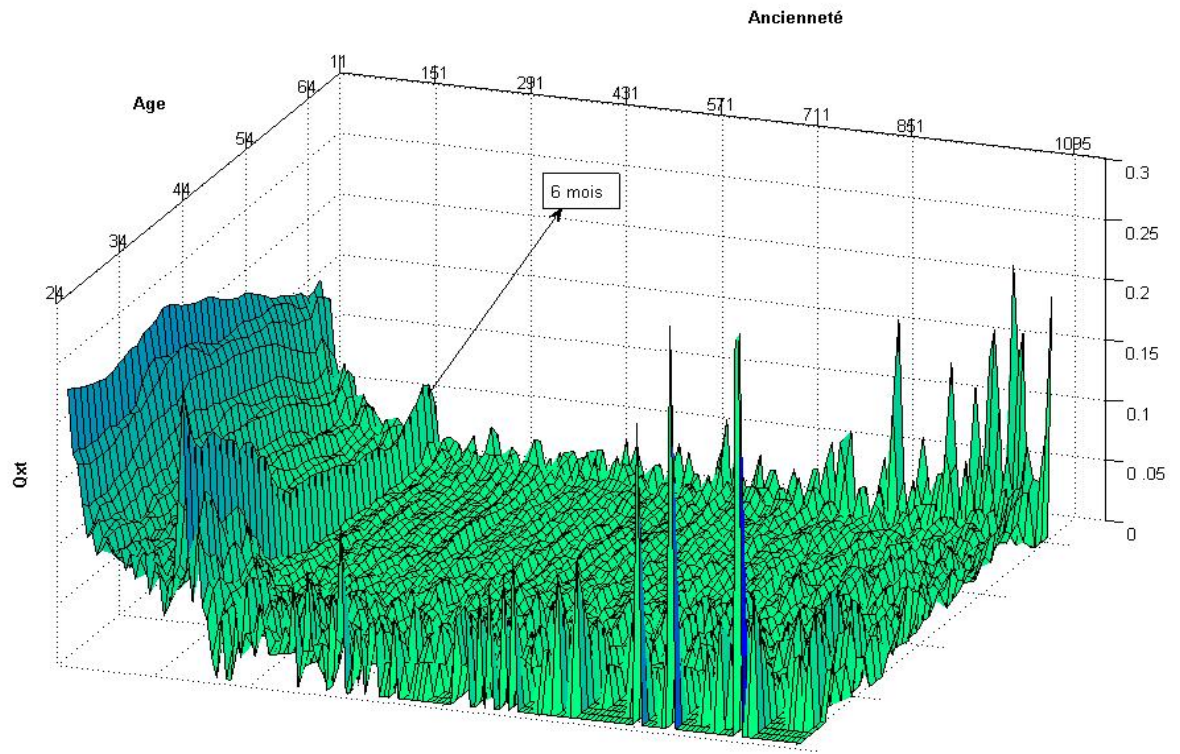

Fig. 2 :  *Taux de sortie d'incapacité bruts (Kaplan-Meier)*

On observe que les taux bruts sont très erratiques, le choix d'une unité de temps petite (le jour) accentuant encore cette variabilité.

Une analyse de la variance donnée par l'estimateur de Greenwood[2] nous permet de conclure que les âges 61, 62, 63 et 64 sont très volatils, les âges 22, 23, 24 sont très volatils, surtout lorsque l'ancienneté est élevée et que la croissance du logarithme de la variance est constante pour les grands âges, alors que sur les âges faibles, la courbe croit beaucoup au début pour ensuite se stabiliser :

---

[2] L'expression utilisée est $Var\hat{S}_x(t) = \hat{S}_x^2(t) \sum_{i=AncMin}^{t} \frac{d_x(i)}{n_x(i)(n_x(i) - d_x(i))}$.

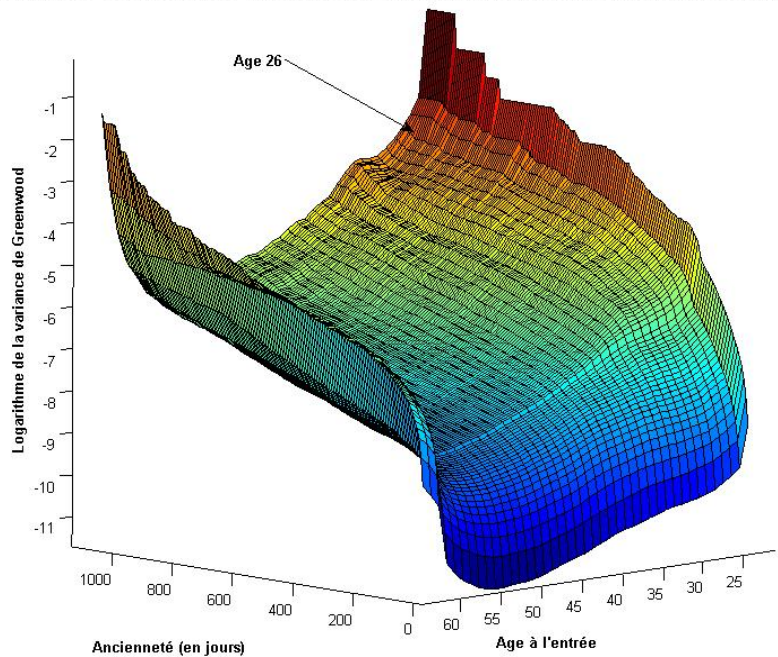

Fig. 3 :    *Estimateur de Greenwood de la variance*

On considérera donc par la suite les âges d'entrée allant de 25 jusqu'à 60 ans, on considérant que ceux-ci sont stables. Les taux de sortie pour les âges en dehors de cette plage pourront être ensuite déterminés par extrapolation (*cf. infra*).

## 3.3. Ajustement par splines

### 3.3.1. Présentation des résultats

Il faut noter ici que si les calculs ne posent pas de problème numérique particulier, ils dépassent les capacités d'un simple outil bureautique tel qu'Excel et nécessitent le recours à des logiciels plus performants. La détermination des paramètres de l'ajustement a été effectuée dans le présent travail en utilisant la librairie « spline » proposée par MatLab. Cette librairie s'appuie sur les algorithmes présentés dans DE BOOR [1978].

La matrice des poids $\Omega$ est construite à l'aide des effectifs sous risques initiaux : pour chaque âge, le coefficient de pondération est proportionnel à l'effectif. La fonction $\lambda$ est prise constante égale à l'unité. Enfin, le coefficient $\alpha$ vaut 0.001. Ce coefficient a été déterminé expérimentalement après différentes tentative. Il apparaît comme le meilleurs compromis entre fidélité et régularité sur la base de nos données.

On obtient ainsi les taux suivants :

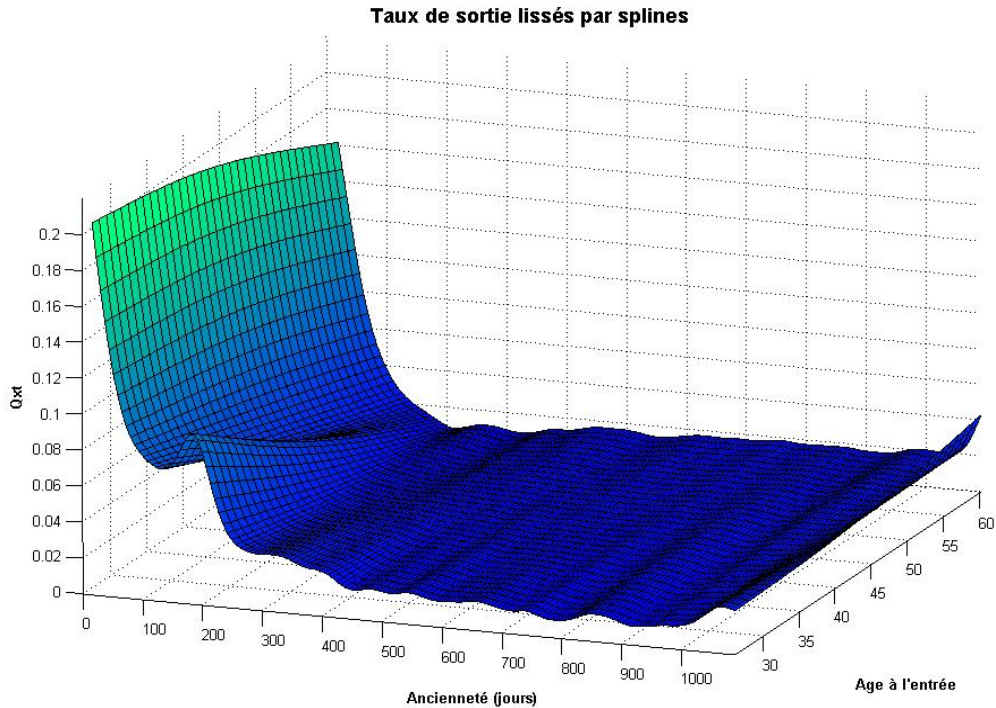

Fig. 4 : *Taux de sortie d'incapacité en fonction de l'âge à l'entrée et de l'ancienneté*

On constate notamment à l'examen du graphe ci-dessus une « pliure » à 6 mois, ainsi que des vaguelettes pour certaines anciennetés ; ceci est la conséquence que les durées des arrêts ne prennent en pratique pas toutes les valeurs possibles exprimées en jours, les prescripteurs d'arrêt ayant tendance à arrondir les durées dans une unité plus en phase avec la durée de l'arrêt (la semaine par exemple).

La surface est bien entendu très régulière : elle est par construction de classe $C^2$. Son expression paramétrique est relativement lourde, mais reste manipulable simplement dans des outils informatiques, par exemple pour calculer des provisions.

### 3.3.2. Test de la validité du modèle

La validité du modèle est testée au travers d'une version *ad hoc* du test du Chi2.

On effectue des regroupements sur les âges et sur les anciennetés, pour obtenir des classes bidimensionnelles. Le test du Chi2 reste valable dans ce contexte puisque la somme de deux variables indépendantes avec une distribution du Chi2 est une variable Chi2, de degré de liberté la somme des degrés de liberté. On obtient alors la statistique :

$$W_{k,l} = \sum_{x^*=1}^{l} \sum_{t^*=1}^{k} \frac{(D_{x^*,t^*} - \tilde{D}_{x^*,t^*})^2}{\tilde{D}_{x^*,t^*}} \qquad (17)$$

avec $D_{x,t}$ le nombre observé d'individus sortis entre $t^+$ et $t+1$ et $\tilde{D}_{x,t}$ le nombre d'individus sortis entre $t^+$ et $t+1$ prévu par le modèle. On peut montrer que si $L_{x,0} \to \infty$, alors $W_{k,l}$ est asymptotiquement distribué comme une variable $\chi^2_{k+l-2}$. D'où le test consistant à rejeter l'adéquation des données brutes à notre ajustement si la réalisation de $W_{k,l}$ est trop grande, c'est-à-dire supérieure à une valeur qui n'a qu'une probabilité $\alpha$ d'être dépassée.

En pratique, nous avons regroupé les âges en neuf classes s'échelonnant de 26 à 60 ans. Quinze classes ont étés constituées sur les anciennetés, ce qui nous donne au total 9+15 = 24 classes. La distribution de la statistique $W_{k,l}$ est donc un Chi2 avec 24 degrés de liberté.

On obtient alors la région critique pour un seuil de 95% $\{W_{k,l} > 13.84\}$ ; la valeur de la statistique de test sur notre exemple est de 10,27, ce qui nous conduit à accepter l'ajustement.

### 3.4. Comparaison avec d'autres méthodes

Il est apparu intéressant de comparer l'approche par splines avec la méthode (non paramétrique) de Whittaker-Henderson, qui constitue, avec la méthode proposée dans KIMELDORF et JONES [1967], une version simple et assez naturelle de lissage bayésien (on pourra consulter sur ce sujet TAYLOR [1992]). La méthode de Whittaker-Henderson appliquée sur les mêmes données conduit à la représentation suivante des taux de sortie[3] :

---

[3] Voir WINTER [2005] pour le détail de la mise en œuvre.

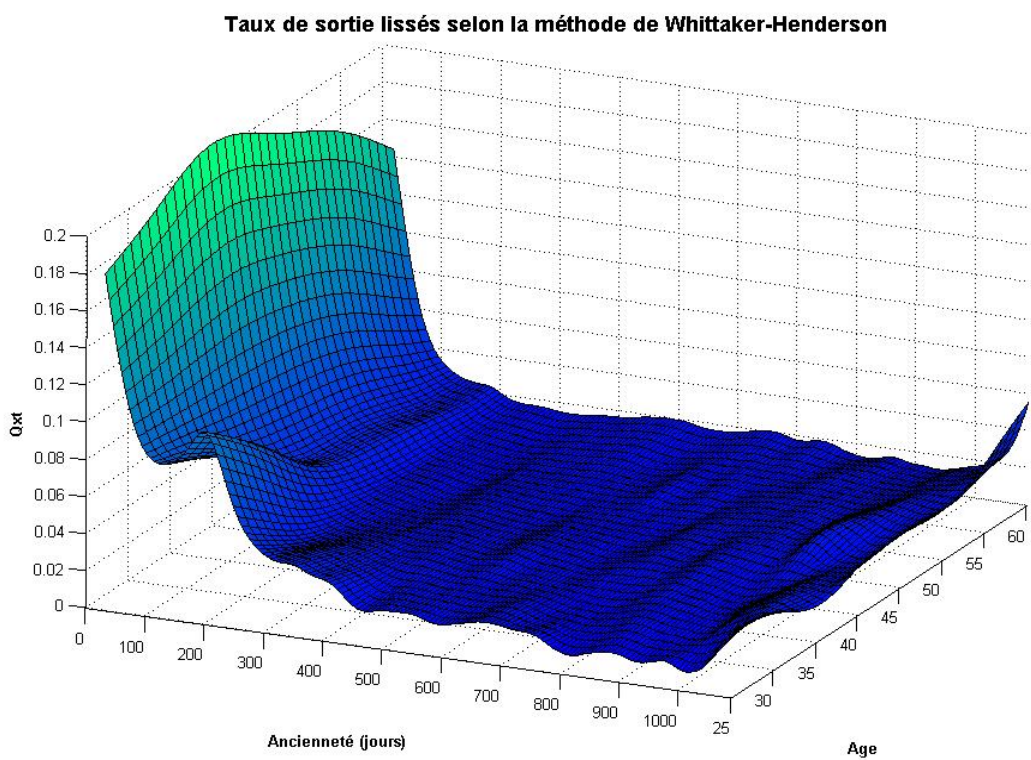

Fig. 5 : *Taux de sortie d'incapacité – lissage de Whittaker-Henderson*

La statistique du Chi2 introduite en 3.3.2 ci-dessus est égal à 6,62, soit une valeur inférieure à celle obtenue par les splines, et qui conduit donc également à accepter l'ajustement au même seuil.

On note d'ailleurs une forte parenté entre les deux approches, l'allure générale des courbes étant proche. De plus chacun des deux modèles permet de combiner un critère de régularité et un critère de fidélité aux données. L'approche par splines présente l'avantage, non négligeable en pratique, de conduire à une formulation entièrement paramétrique de la table, ce qui permet de traiter de manière simple les questions d'interpolation ou d'extrapolation. L'extrapolation des taux de sortie est en effet décisive pour compléter la table aux âges où les données sont insuffisantes. Pour cette raison, elle doit à notre sens être préférée à l'approche non paramétrique.

WINTER [2005] développe également d'autres approches, qui ne sont pas reprises ici ; on peut simplement retenir que du point de vue de critères de type « chi2 », la méthode d'ajustement par splines s'avère très performante.

## 3.5. Une approche alternative

On peut observer que l'approche retenue conduit à des coefficients de provisionnement très réguliers : en effet, ces coefficients sont obtenus par deux intégrations successives à partir des taux de sortie, qui sont ajustés par des splines (une première intégration pour obtenir la fonction de survie, puis une seconde pour calculer le coefficient de provisionnement). De ce fait, même si nous étions parti de taux de sortie relativement irréguliers, les coefficients de provisionnement seraient réguliers (les tables du BCAC en sont une illustration, les taux de sortie étant très irréguliers).

On peut envisager d'exploiter cette propriété de régularisation du mécanisme de calcul des provisions pour proposer la démarche d'ajustement suivante : à partir des taux bruts, on détermine les espérances de maintien résiduel (qui s'interprètent comme des coefficients de provisionnement à taux zéro) :

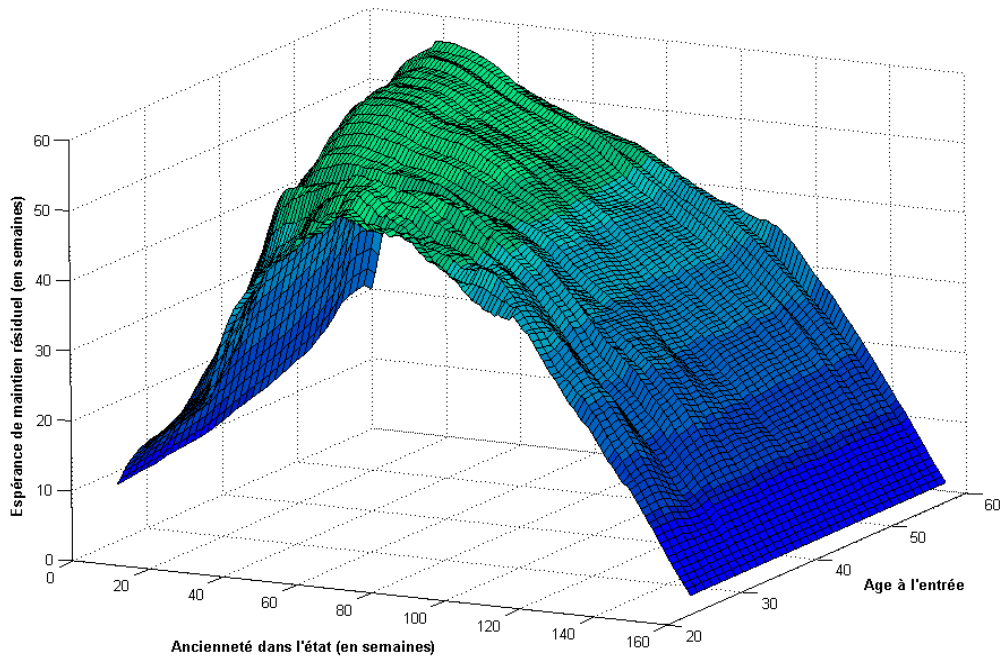

Fig. 6 :     *Espérance de maintien résiduel brute*

On note bien l'effet de régularisation associé au calcul d'un coefficient de provisionnement, en rapprochant ce graphe de celui obtenue à la Fig. 2 : ci-dessus. On ajuste alors un modèle à splines sur cette surface afin de la régulariser, ce qui conduit à :

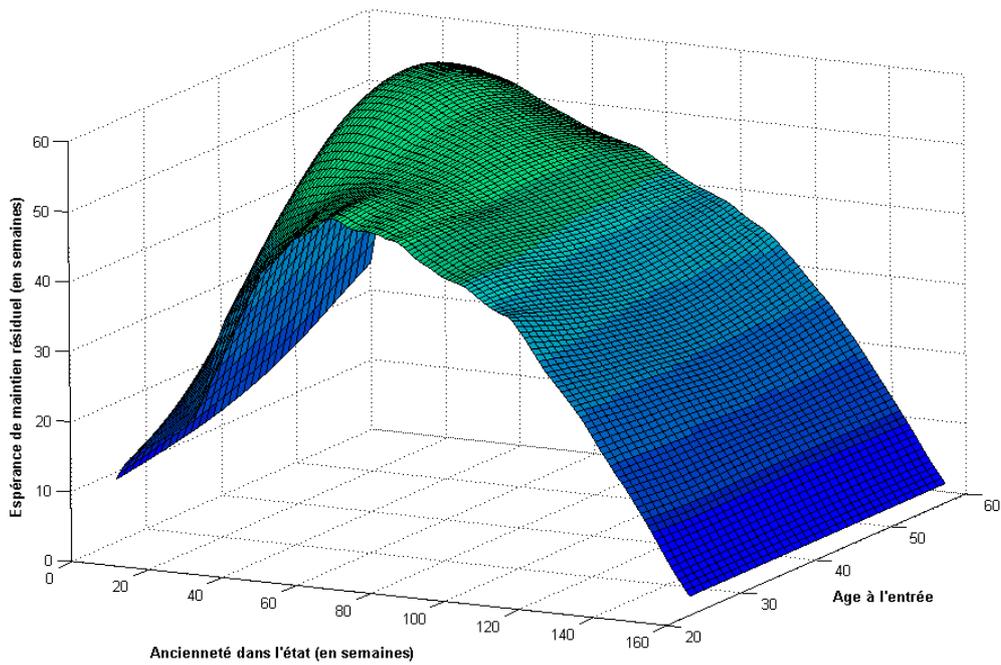

Fig. 7 :   *Espérance de maintien résiduel lissée*

On peut ensuite « redescendre » aux taux de sortie par dérivation. On obtient alors les taux de sortie suivants :

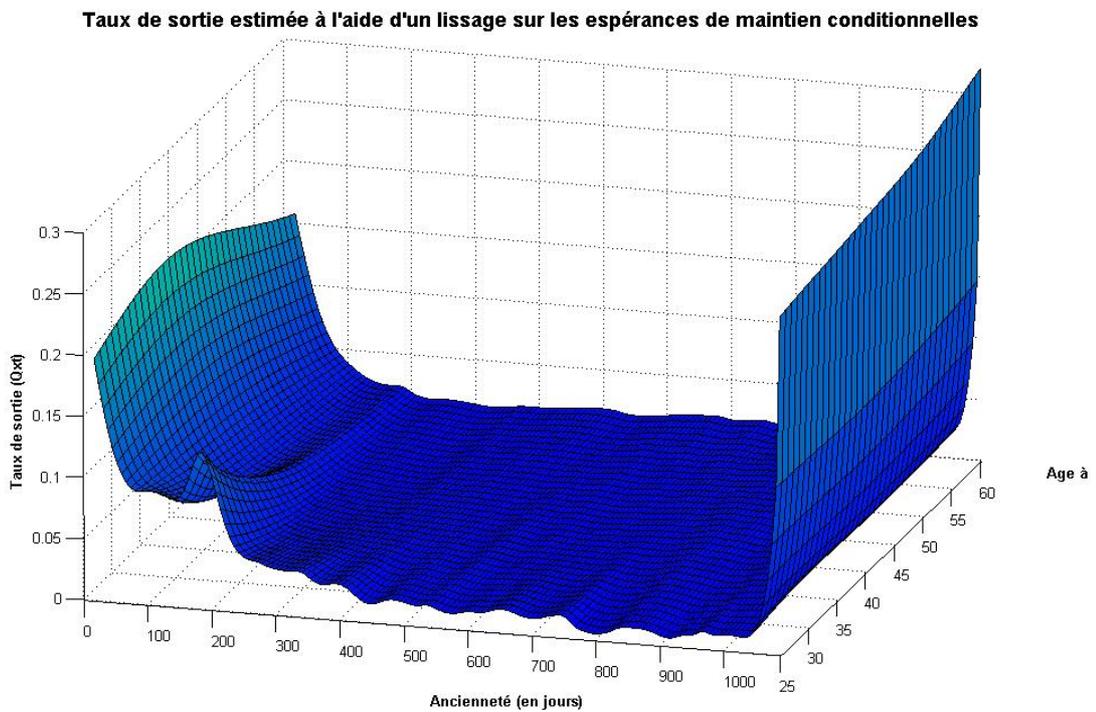

Fig. 8 :   *Taux de sortie après ajustement spline des espérances de maintien*

L'allure générale de la surface est relativement proche des résultats précédents, et la valeur de la statistique de test du même ordre de grandeur. Toutefois, il nous semble qu'il est plus pertinent d'effectuer l'ajustement au niveau le « moins intégré » possible, et donc au niveau des taux de sortie.

## 4. Application : calcul d'un barème de provisionnement

L'utilisation la plus immédiate de la table d'expérience construite ci-dessus est l'obtention d'un barème de provisionnement de rentes d'incapacité. Pour un taux technique $i$ fixé, le coefficient de provisionnement s'exprime de la manière suivante :

$$_{INC}PM_y^x = \frac{1}{_{INC}L_y^x} \sum_{k=0}^{36-y} \frac{_{INC}L_{k+y}^x}{(1+i)^{\frac{k}{12}}}, \qquad (18)$$

avec $x$ l'âge de l'assuré lors de l'entrée dans l'état, $y$ l'ancienneté dans l'état et $_{INC}L_y^x$ le coefficient de la table de maintien en incapacité.

Incidemment, on peut noter que dans le cadre du modèle retenu il n'est pas indispensable de discrétiser les flux, ce qui permet d'utiliser une version continue de la formule ci-dessus :

$$_{INC}PM_y^x = \frac{1}{_{INC}L_y^x} \int_0^{+\infty} {_{INC}L_{t+y}^x} \exp(-rt)dt, \qquad (19)$$

avec $r = \ln(1+i)$ la version continue du taux technique et $_{INC}L_{t+y}^x = \exp\left(-\int_0^{t+y} \mu_x(\theta)d\theta\right)$.

Cette formulation présente l'avantage d'éviter les problématiques de choix d'un pas de discrétisation, d'une part, et d'ajustement en fonction du fractionnement, d'autre part. Dans l'expression ci-dessus, les paramètres $x$ et $y$ peuvent être non entiers.

Le barème ainsi obtenu a l'allure suivante, avec un taux technique de 3% :

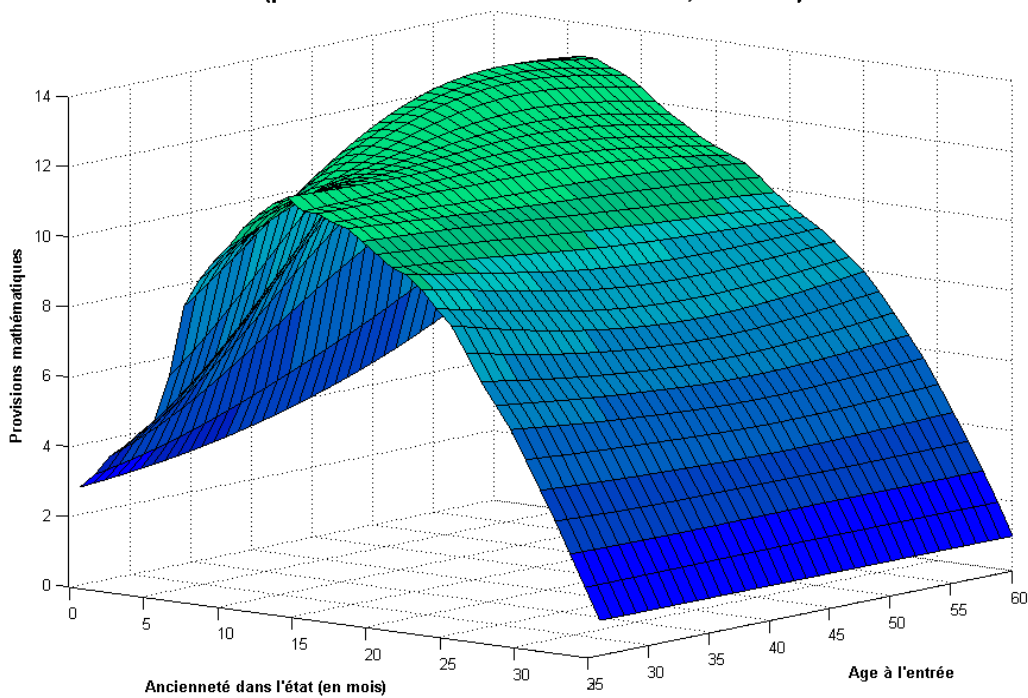

Fig. 9 :        *Barème de provisionnement d'incapacité d'expérience*

On retrouve bien entendu une allure très proche de la Fig. 7 : ci-dessus. Le barème du BCAC au même taux conduit quant à lui à la surface suivante :

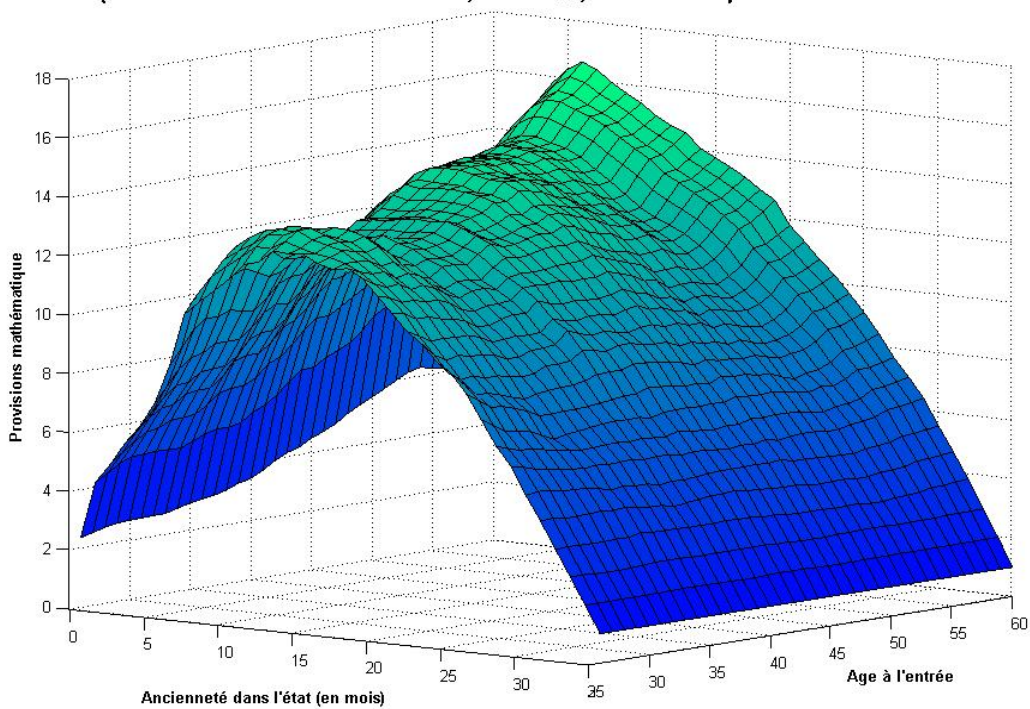

Fig. 10 :       *Barème de provisionnement d'incapacité BCAC*

Incidemment, on note une certaine irrégularité du barème BCAC, conséquence directe de l'utilisation de taux de sortie non ajustés dans la démarche retenue par cet organisme. La forme générale des deux barèmes est toutefois, comme on pouvait s'y attendre, très proche. La comparaison des deux séries de coefficients est effectuée en calculant
$\forall (x,t) \quad \dfrac{e_{x,t}^{BCAC} - e_{x,t}^{Splines}}{e_{x,t}^{Splines}}$ ; elle conduit à la représentation suivante :

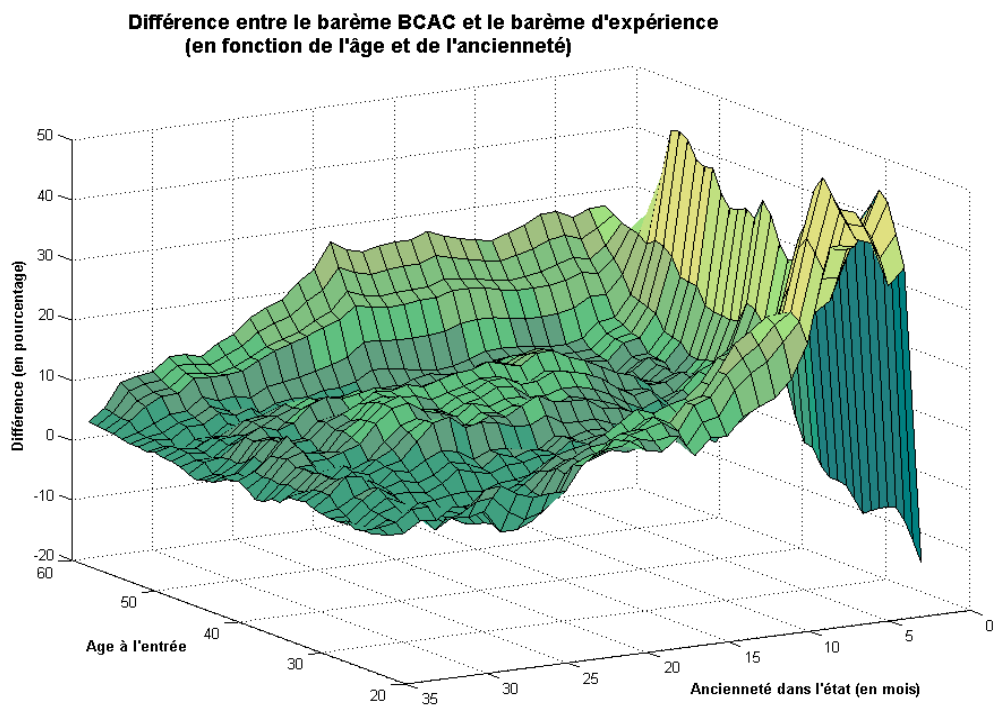

Fig. 11 : *Comparaison avec les coefficients BCAC*

On constate que globalement le barème BCAC conduit à sur-provisionner d'environ 5% à 10% en milieu de table. Pour les anciennetés importantes, les deux tables donnent des résultats proches. Par contre, au moment de la survenance, les résultats d'expérience sont au dessus du BCAC : l'effectif restant après le premier mois varie de 30% à 50% pour le BCAC, et de 50% à 70% pour les taux d'expérience. Ensuite, pour une ancienneté de 1 mois, le BCAC conduit à sur provisionner de 30% à 40%. L'écoulement des sorties est donc sensiblement différent dans les deux populations, ce qui justifie *ex post* l'intérêt du barème d'expérience.

Le graphique ci-dessus met de plus en évidence l'instabilité du barème du BCAC, qui explique les irrégularités que l'on observe (la surface d'expérience étant quant à elle par construction régulière). L'avantage de la méthode d'ajustement retenue est de conduire mécaniquement à des barèmes très réguliers, *a priori* plus en phase avec la réalité.

# 5. Conclusion

L'utilisation d'un ajustement des taux bruts par des splines bi-dimensionnels permet de construire une surface de taux de sortie paramétrique. Sa mise en œuvre numérique ne pose pas de difficulté majeure avec des outils adaptés.

L'intérêt de cette approche, par rapport aux méthodes usuelles, repose principalement sur la prise en compte directe dans le modèle de la dépendance entre les lois de maintien aux différents âges d'entrée possibles.

On obtient ainsi une prise en compte directe des interactions dans les deux directions (âge et ancienneté).

Au surplus, cette approche bi-dimensionnelle permet d'obtenir des lois de maintien pour des âges à l'entrée pour lesquels les observations sont insuffisantes pour construire directement la loi d'expérience, par simple interpolation.

Enfin, le caractère paramétrique du modèle permet de disposer de valeurs des taux de sortie quelle que soit l'unité de temps retenue, ce qui simplifie les calculs de provisions, quel que soit le fractionnement de la rente.

La mise en œuvre des techniques présentées ici dans le cas de l'invalidité est immédiat ; des travaux sont en cours dans le prolongement de cette étude pour construire, à partir de la surface ajustée des taux de maintien conditionnels[4], une surface « au premier jour », avec l'objectif de disposer ainsi d'un outil permettant de tarifer des contrats individuels qui prendraient en charge les premiers jours de l'arrêt. Nous nous inspirons pour cela de la démarche proposée par BONCHE et al. [2005].
.

---

[4] On rappelle que la durée d'arrêt minimale dans l'étude est de 11 jours.

# Bibliographie